\pdfoutput=1

\documentclass[superscriptaddress,prx,twocolumn,nofootinbib,notitlepage]{revtex4-1}

\usepackage{graphics,graphicx,amsthm,amsmath,amssymb}
\usepackage{color}

\def\({\left(}
\def\){\right)}

\def\bra#1{\mathinner{\langle{#1}|}}
\def\ket#1{\mathinner{|{#1}\rangle}}
\def\braket#1#2{\mathinner{\langle{#1}|#2 \rangle}}

\def\avg#1{\mathinner{\langle{#1} \rangle}}
\def\var#1{\mathinner{\rm Var\({#1}\)}}

\def\IQP{{\rm \scriptscriptstyle IQP}}
\def\PT{{\rm \scriptscriptstyle PT}}
\def\PR{{\rm \scriptscriptstyle PR}}
\def\S{{\rm \scriptscriptstyle S}}
\def\cl{{\rm \scriptscriptstyle pcl}}
\def\F{{\rm \scriptscriptstyle F}}
\def\H{{\rm H}}
\def\mE{{\mathbb E}}

\begin{document}

\title{Fourier analysis of sampling from noisy chaotic quantum circuits.}

\author{Sergio Boixo}
\affiliation{Google Inc., Venice, CA 90291, USA}
\author{Vadim N. Smelyanskiy}
\affiliation{Google Inc., Venice, CA 90291, USA}
\author{Hartmut Neven}
\affiliation{Google Inc., Venice, CA 90291, USA}

\date{\today}

\begin{abstract}
  Sampling from the output distribution of chaotic quantum evolutions, and of pseudo-random universal quantum circuits in particular, has been proposed as a prominent milestone for near-term quantum supremacy. The same paper notes that chaotic distributions are very sensitive to noise, and under quite general noise models converge to the uniform distribution over bit-strings exponentially in the number of gates. On the one hand, for increasing number of gates, it suffices to choose bit-strings at random to approximate the noisy distribution with fixed statistical distance. On the other hand, cross-entropy benchmarking can be used to gauge the fidelity of an experiment, and the distance to the uniform distribution. We estimate that state-of-the-art classical supercomputers would fail to simulate high-fidelity chaotic quantum circuits with approximately fifty qubits and depth forty. A recent interesting paper proposed a different approximation algorithm to a noisy distribution, extending previous results on the Fourier analysis of commuting quantum circuits. Using the statistical properties of the Porter-Thomas distribution, we show that this new approximation algorithm does not improve random guessing, in polynomial time. Therefore, it confirms previous results and does not represent an additional challenge to the suggested failure stated above. 

\end{abstract}
\maketitle

We are entering a period of time where experimental quantum devices of growing size and fidelity will perform well defined computational tasks, of progressing practical value, beyond the capabilities of state-of-the-art classical supercomputers. Sampling problems, where the goal is to approximately sample from a well defined probability distribution, are prominent examples of such tasks, and also possess interesting relations to computational complexity theory~\cite{aaronson2011computational,bremner2015average,boixo2016characterizing,bremner16,gao2017quantum}. 

Reference~\cite{boixo2016characterizing} proposed sampling from  a chaotic quantum evolution as a particularly promising computational task for a near-term ``quantum supremacy''~\cite{preskill_2012} demonstration. More specifically, the task is to approximately sample from the output distribution $\{p_U(x)\}$ (over bit-strings $\{x\}$) of a pseudo-random universal quantum circuit $U$. We are particularly interested in 2D circuits of depth $O(\sqrt n)$, where $n$ is the number of qubits. 
Classically sampling from a distribution correlated with the output distribution of $U$ likely requires exponential classical resources (when averaged over instances in an ensemble). This can be argued in two different ways. From a physical point of view, pseudo-random universal quantum circuits, as examples of chaotic quantum evolutions, are hypersensitive to perturbations~\cite{dahlsten2007emergence,scott_hypersensitivity_2006}. Furthermore, quantum states from instances of an ensemble of chaotic evolutions spread quasi-uniformly in Hilbert space, and each output distribution quickly approaches the entropy and lower moments of the characteristic Porter-Thomas (or exponential) distribution~\cite{porter1956fluctuations,schack_hypersensitivity_1993,beenakker1997random,emerson2003pseudo,dahlsten2007emergence,trail_entanglement_2008,harrow2009random,weinstein_parameters_2008,brown2010convergence,brown_scrambling_2012,PhysRevLett.111.127205,hosur2015chaos,boixo2016characterizing}.  Therefore, sampling from the output distribution is expected to require a direct simulation of the quantum dynamics, which classically implies exponential cost. The same paper~\cite{boixo2016characterizing} also extends previous results in computational complexity theory~\cite{aaronson2003quantum,terhal2004adaptive,aaronson2005quantum,bremner_classical_2011,aaronson2011computational,fujii2013quantum,aaronson2014equivalence,fujii2014impossibility,jozsa_classical_2014,bremner2015average}, and in particular of Ref.~\cite{bremner2015average}, to argue the same point. 

More specifically, assume a polynomial classical polynomial algorithm which takes a description of circuit $U$ as input. We write its output probabilities as  $\{p_\cl(x|U)\}$. Our assumption is that there is no polynomial time classical algorithm for which the average cross-entropy between $\{p_\cl(x|U)\}$ and the ideal probabilities $\{p_U(x)\}$ is less than~\cite{boixo2016characterizing}
\begin{multline}
  \mE_U \left[ \sum_{j=1}^N p_\cl(x_j|U) \log \frac 1 {p_U(x_j)} \right] \\ \le \log N + \gamma - \Omega\( \frac 1 {N}\)\;,
\end{multline}
where $N=2^n$ and $\gamma$ is the Euler's constant.
The value $\log N + \gamma$ corresponds to the case where $p_\cl(x|U)$ is uncorrelated with $p_U(x)$, assuming enough depth so that the entropy  of $\{p_U(x)\}$ is well approximated by the entropy from the Porter-Thomas distribution $\mE_U [\H(p_U)] = \log N + \gamma -1 $, with $U$ dependent fluctuations of order $2^{-n/2}$. The reason why we focus on the cross entropy is because it is approximately linear in the circuit fidelity. We use this to estimate the fidelity using cross-entropy benchmarking. 
A related quantum threshold conjecture was presented recently, stating that there is no polynomial classical algorithm which can guess if $p_U(x)$ is above or below the median with probability at least $1/2 + \Omega(1/N)$~\cite{aaronson2016complexity}~\footnote{It is formally possible that an algorithm could brake the cross-entropy assumption without braking the quantum threshold conjecture~\cite{aaronson2016complexity}.}. 

Nevertheless, under fairly general noise models, the output distribution of chaotic quantum dynamics converges to the classical uniform distribution~\cite{boixo2016characterizing}. Therefore, in the asymptotic limit, it suffices to sample uniformly at random to approximate the experimental output for any fixed distance $\delta$ in the $\ell_1$ norm. This can be done classically in linear cost on the number of qubits $n$. In other words, ``achieving a constant error in the limit of large $n$ requires a fault tolerant quantum computer, which will not be available in the near term~\cite{kalai2014gaussian,arkhipov2015bosonsampling,rahimi2016sufficient}''~\cite{boixo2016characterizing}. It is therefore critical for experimentally accessible near-term quantum supremacy experiments to count with a well-defined metric for the relevant computational task. In this vein, one of the main contributions of Ref.~\cite{boixo2016characterizing} is the introduction of cross-entropy benchmarking as an approximation to the fidelity, and to the distance to the uniform distribution, for complex quantum systems.


Reference~\cite{boixo2016characterizing} also gives an ansatz for the density matrix produced by a noisy chaotic quantum evolution
\begin{align}
  \rho \simeq   \alpha \ket{\psi}\bra{\psi} + (1-\alpha) \frac \openone {N}\;,\label{eq:rho}
\end{align}
where $\ket \psi$ is the ideal output and $\alpha$ is the fidelity. This ansatz is supported numerically by the observation that the output distribution after adding a single discrete error is (almost) uncorrelated with the ideal output distribution, and noticing that $\alpha$ is (almost) the probability of no-error in the quantum circuit. It is also consistent with simulations where each ideal gate is followed by a depolarizing channel with error rate $\epsilon$, a common error model well matched by experimental results~\cite{barends_superconducting_2014,barends_digital_2015,kelly_state_2015,barends_digitized_2015,emerson_scalable_2005,knill2008randomized,magesan_robust_2011,magesan_characterizing_2012}. 
The fidelity $\alpha$ can  be approximated by $\alpha \simeq e^{- \epsilon m}$, where $m$ is the number of gates and $\epsilon$ is the error per gate. Therefore, classically it suffices to choose bit-strings $x$ uniformly at random to approximate the output distribution of a noisy quantum circuit with statistical distance $\delta \simeq e^{-\epsilon m}$. We note also that noisy quantum circuits  do not violate asymptotically the cross-entropy assumption or the quantum threshold conjecture stated above for polynomial classical algorithms. 

Nonetheless, a quantum computer with gate error rates within reach in the near term would be able to approximate the ideal distribution with an statistical distance beyond the capabilities of state-of-the-art classical supercomputers. For instance, two-qubit, initialization and measurement error rates of 0.3\%, with single-qubit error rates of 0.06\%, would result in a final fidelity of approximately 10\% for circuits of $7\times7$ qubits in a 2D lattice and depth 40. This depth takes into account current constraints in the layout of two qubit gates for superconducting qubits~\cite{barends_superconducting_2014,barends_digital_2015,kelly_state_2015,barends_digitized_2015,boixo2016characterizing}. Sampling a correlated distribution classically would require $2^{49} \times 8 \times 2$ bytes in Rapid Access Memory, which is likely to be unfeasible. The conjectures stated above give support to the estimated exponential cost of classical approximation algorithms. 

\begin{figure}
  \centering
  \includegraphics[width=\columnwidth]{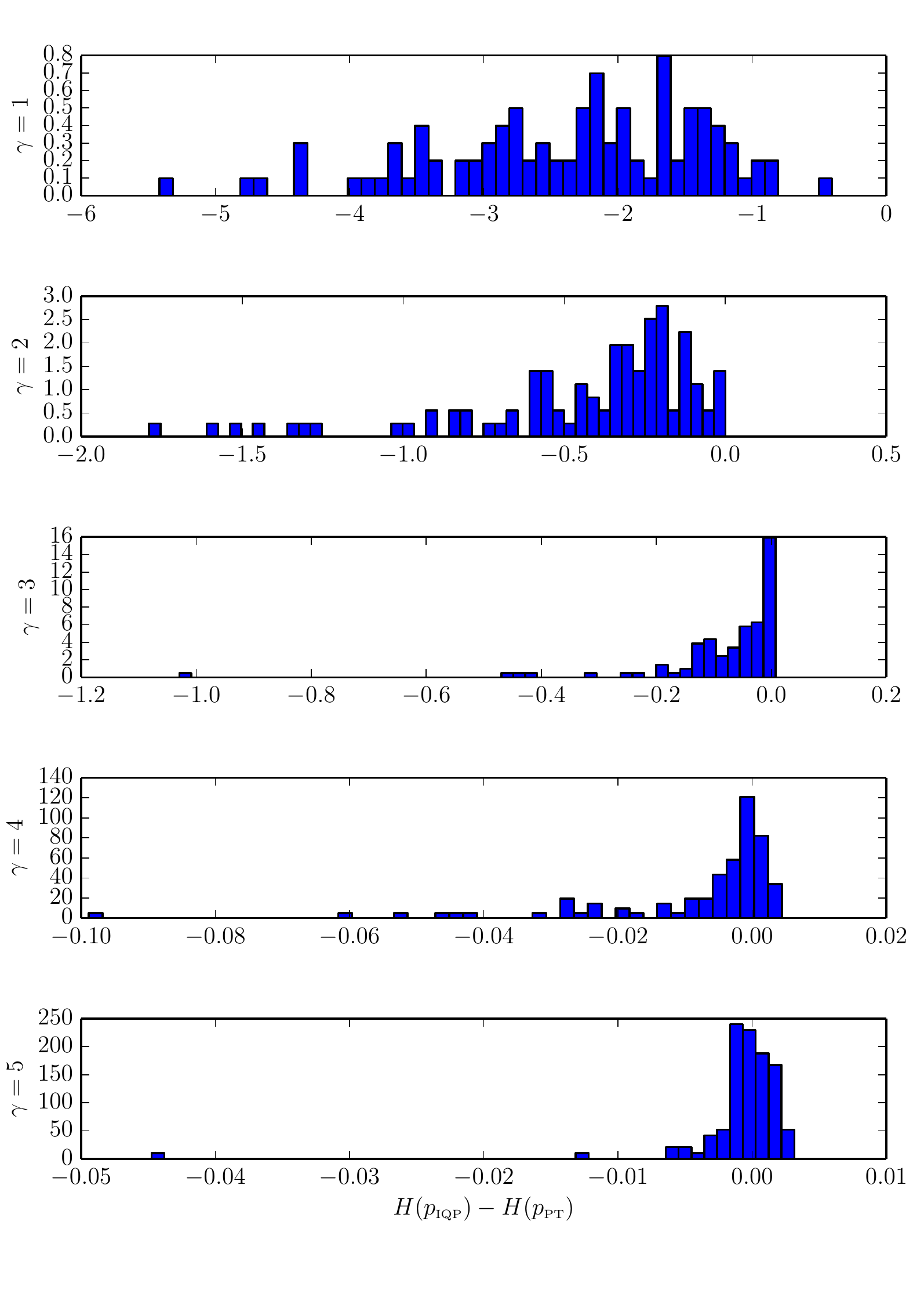}
  \caption{Histograms of the entropy  $H(p_\IQP)$ of the output distribution of sparse IQP circuits~\cite{bremner16} with 20 qubits showing the convergence to the entropy of the corresponding Porter-Thomas distribution $ H(p_{\rm \scriptscriptstyle PT})$ with increasing circuit density, controlled by $\gamma$. For each possible choice of a pair $(j,k)$ of distinct qubits, we apply a controlled-phase gate across those qubits with probability $\gamma(\log n)/n$. }
  \label{fig:entropy_vs_gamma}
\end{figure}

\begin{figure}
  \centering
  \includegraphics[width=\columnwidth]{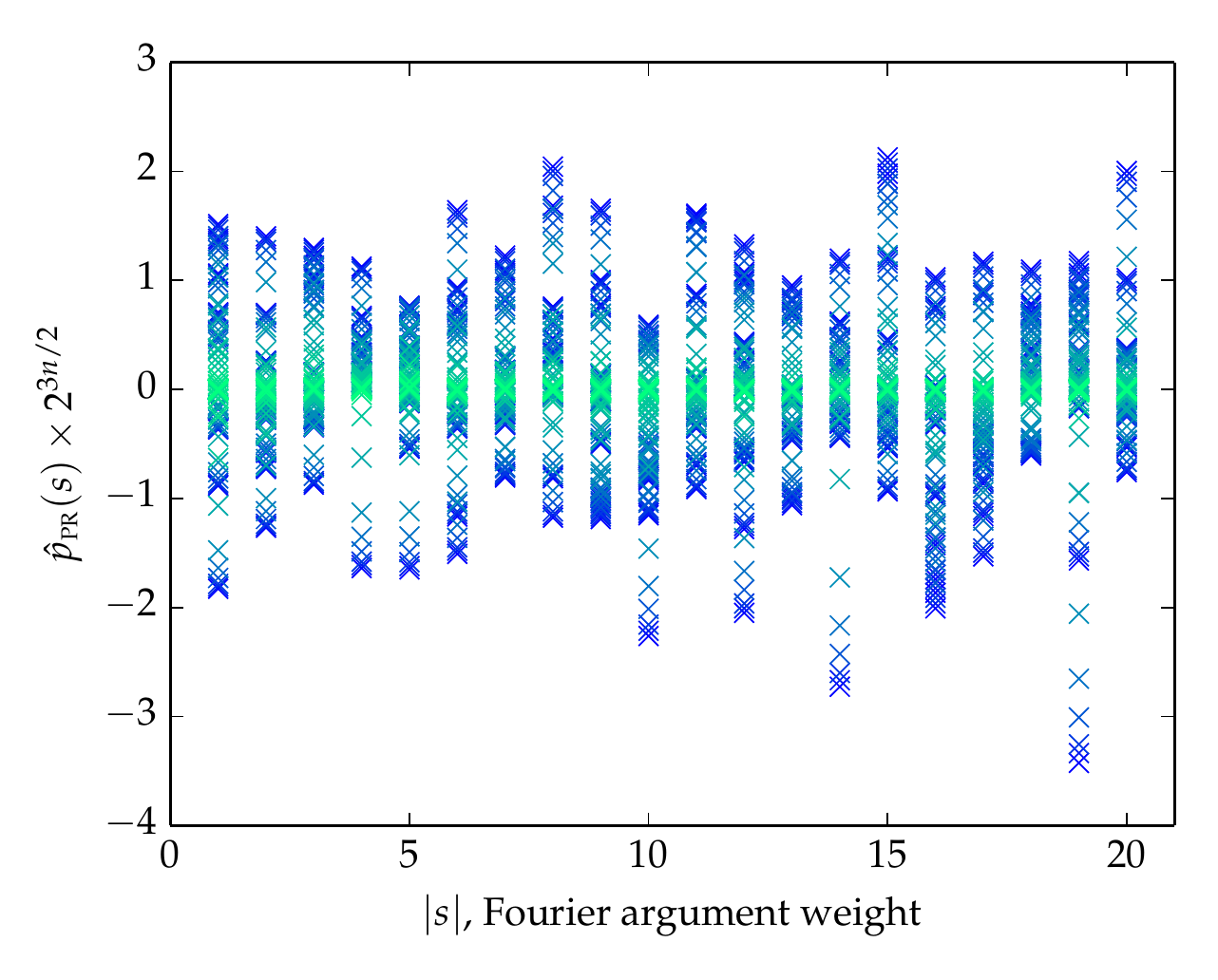}
  \caption{Numerical Fourier components for the output distribution of pseudo-random universal quantum circuits~\cite{boixo2016characterizing} with $5 \times 4$ qubits and depth 40, rescaled  by the standard deviation $2^{-3 n /2}$ from Porter-Thomas. The two-qubit gate error rate is, from blue to green, $\epsilon =  [0, 0.0001, 0.0002, 0.0005, 0.001, 0.002, 0.005, 0.01, 0.02, 0.05]$,  the single-qubit gate error rate is 10 times less. We plot 10 instances and 10 Fourier components for each error rate and weight $l = |s|$. Numerics confirm the Porter-Thomas estimates, and all components converge to 0 (uniform random distribution) with increased noise. }
  \label{fig:fourier_component}
\end{figure}

A recent interesting paper~\cite{yung2017can} extends results on commuting or IQP quantum circuits~\cite{bremner16} to universal quantum circuits, and obtains a polynomial time classical algorithm for approximating the output distribution under certain models of noise. The upper bound on the cost is $(n+m)^{O(\frac 1 \epsilon \log \frac 1 \delta)}$ where $n$ is the number qubits, $m$ the number of gates, $\epsilon$ the error per gate, and $\delta$ the desired distance in the $\ell_1$ norm from the noisy distribution. 
A first observation is that for the experimental error rates $\epsilon \simeq 0.3 \%$ within reach of experimental quantum computers, the estimate of this upper bound is much worse than a direct simulation for sizes of practical interest~\cite{bremner16}. Furthermore, the upper bound to improve over the statistical distance $\delta \simeq e^{- \epsilon m}$ achieved by random sampling of bit-strings is $(n+m)^{O(m)}$, again worse than exact simulation. We next analyze this algorithm in more detail and conclude that indeed, when restricted to polynomial time, it results in an output distribution whose correlation with the ideal distribution is exponentially small in the number of gates $m \gg n$ for chaotic quantum circuits.

The methodology of Ref.~\cite{yung2017can} proceeds in two steps. First, a universal random circuit is mapped to an IQP circuit using techniques related to measurement-based quantum computation. The resulting IQP circuit has $n'=n+m$ qubits. Second, the noisy output distribution is approximated using the Fourier analysis algorithm from Ref.~\cite{bremner16}. For our purposes it will suffice to focus in the second step.

Let's now review the algorithm from Ref.~\cite{bremner16}. Consider an IQP circuit $U_\IQP = H^{\otimes n} D H^{\otimes n}$ acting on $n$ qubits, where $H$ is the Hadamard gate and $D$ is a diagonal circuit in the computational basis, composed of $Z$ rotations and control-$Z$ gates. The output probability of bit-string $x$ is
\begin{align}
  p_\IQP(x) &= |\bra x U_\IQP \ket 0 |^2 \\ &=\left|  \frac 1 {2^{n}}\sum_y f(y) (-1)^{x \cdot y}\right|^2\;,
\end{align}
where
\begin{align}
  f(y) = \bra y D \ket y\;.
\end{align}
Note that $f(y)$ can be classically computed in polynomial time in $n$. 

For any boolean function $f(x) : \{0,1\}^{n} \to \mathbb{C}$, we can define a Fourier transform as
\begin{align}
  \hat f (s) &= \frac 1 {2^{n}} \sum_x f(x)(-1)^{x \cdot s} \\
   f(x) &= \sum_s \hat f(s) (-1)^{x \cdot s} \;.
\end{align}
Therefore
\begin{align}
   p_\IQP(x) = \left| \hat f(x) \right|^2\;,
\end{align}
and by the convolution theorem we have
\begin{align}
  \hat   p_\IQP(s) = \frac 1 {2^{n}} \sum_y f^*(y) f(y+s)\;.
\end{align}
It can be seen that~\cite{bremner16}, from the Chernoff bound, we can approximate $2^{n} \hat p_\IQP(s)$ up to an additive error $\eta$ in time $O(1/\eta^2)$ using the same order of evaluations of $f$. 

Consider now an output distribution $p^\epsilon_\IQP(x)$ where a depolarizing channel with error rate $\epsilon$ is added to each qubit, exactly before measurement. Note that a limitation of this model is that errors do not spread, because there are no errors prior to any gate. The motivation is that this noise model can be studied using tools from Fourier analysis. Consider the Fourier transform $\hat p^\epsilon_\IQP(s)$ of $p^\epsilon_\IQP(x)$. It can be shown that this error model diminishes the Fourier coefficients $\hat p^\epsilon_\IQP(s)$ exponentially in the Hamming weight $l = |s|$ of the argument $s$~\cite{bremner16,o2014analysis}
\begin{align}
  \hat p^\epsilon_\IQP(s) = (1-\epsilon)^{|s|} \hat p_\IQP(s)\;.\label{eq:pes}
\end{align}
Note that the uniform distribution over bit-string has Fourier components $\hat p_{\rm uniform}(s) = 2^{-n}\delta_{s,0}$. Equation~\eqref{eq:pes} then says that the Fourier components of $p_\IQP$ converge to the Fourier components of the uniform distribution exponentially in $|s|$. There is some discrepancy with the ansatz of Eq.~\eqref{eq:rho}, which implies that the Fourier components converge to those of the uniform distribution exponentially in the number of gates $m$, and independently of $|s|$. Nevertheless, for this ansatz we are assuming a very different error model: we apply a depolarizing channel after each gate in a circuit with significant depth, so errors can propagate to most qubits. This is in contrast to the noise model that led to Eq.~\eqref{eq:pes}, where errors are constrained to the qubit in which they occur.

The algorithm in Ref.~\cite{bremner16} consists in approximating a polynomial number $O(n^l)$ of Fourier components with low weight $|s| \le l$, up to polynomial error $O(\delta  n^{-l/2})$. According to Eq.~\eqref{eq:pes}, Fourier components with higher weight are diminished by a factor of at least $(1-\epsilon)^l$. Now assume that $\sum p_\IQP^2(x) \le \beta  2^{-n}$ for some constant $\beta$ independent of $n$. It is then shown that it suffices to choose $l = O(\log \beta/\delta)/\epsilon)$ for a target statistical distance $\delta$. An upper bound  $O(n^{O(\frac 1 \epsilon \log \frac 1 \delta)})$ on the cost of this algorithm follows from a bound on the number of Fourier components $O(n^l)$ being  approximated, and the previous bound on the cost to approximate each component.

We now estimate the distribution of Fourier components. The output probabilities $p_\PT(x) = |\braket x \psi|^2$ from a state $\ket \psi$ chosen uniformly at random in Hilbert space have a characteristic distribution called the Porter-Thomas (or exponential) distribution~\cite{porter1956fluctuations,brody_random-matrix_1981,mehta2004random}. They are i.i.d distributed (up to normalization) with $\Pr(p_\PT) = N e^{-N p_\PT}$. Figure~\ref{fig:entropy_vs_gamma} shows that the entropy of the output distribution of IQP circuits approximates the  Porter-Thomas distribution. This is also true a fortiori for the output probabilities $\{p_\PR(x)\}$ of pseudo-random universal quantum circuits~\cite{boixo2016characterizing}. We therefore will use the approximation
\begin{align}
  \Pr_\IQP \simeq \Pr_\PT \simeq \Pr_\PR
\end{align}
for the distribution of output probabilities. 

Under this approximation to Porter-Thomas we have (for $s \ne 0$)
\begin{align}
  \avg {\hat p_\PT(s) }_{\ket \psi} &=\frac 1 {2^n} \bigg\langle \sum_x p_\PT(x) (-1)^{x\cdot s} \bigg\rangle_{\ket \psi} = 0\;.
\end{align}
That is, the expectation of each Fourier component is the same as for the uniform distribution. The variance is
\begin{align}
  \var  {\hat p_\PT(s) } &=\frac 1 {2^{2n}} \var{ \sum_x p_\PT(x) (-1)^{x\cdot s}} \\
  &= \frac 1 {2^n} \var{p_\PT} = 2^{-3n}\;.
\end{align}
More specifically, we show in App.~\ref{app:pt_fourier_distr} that the distribution of Fourier components of probabilities with a Porter-Thomas distribution is Gaussian with mean 0 and standard deviation $2^{-3n/2}$.

\begin{figure}
  \centering
  \includegraphics[width=\columnwidth]{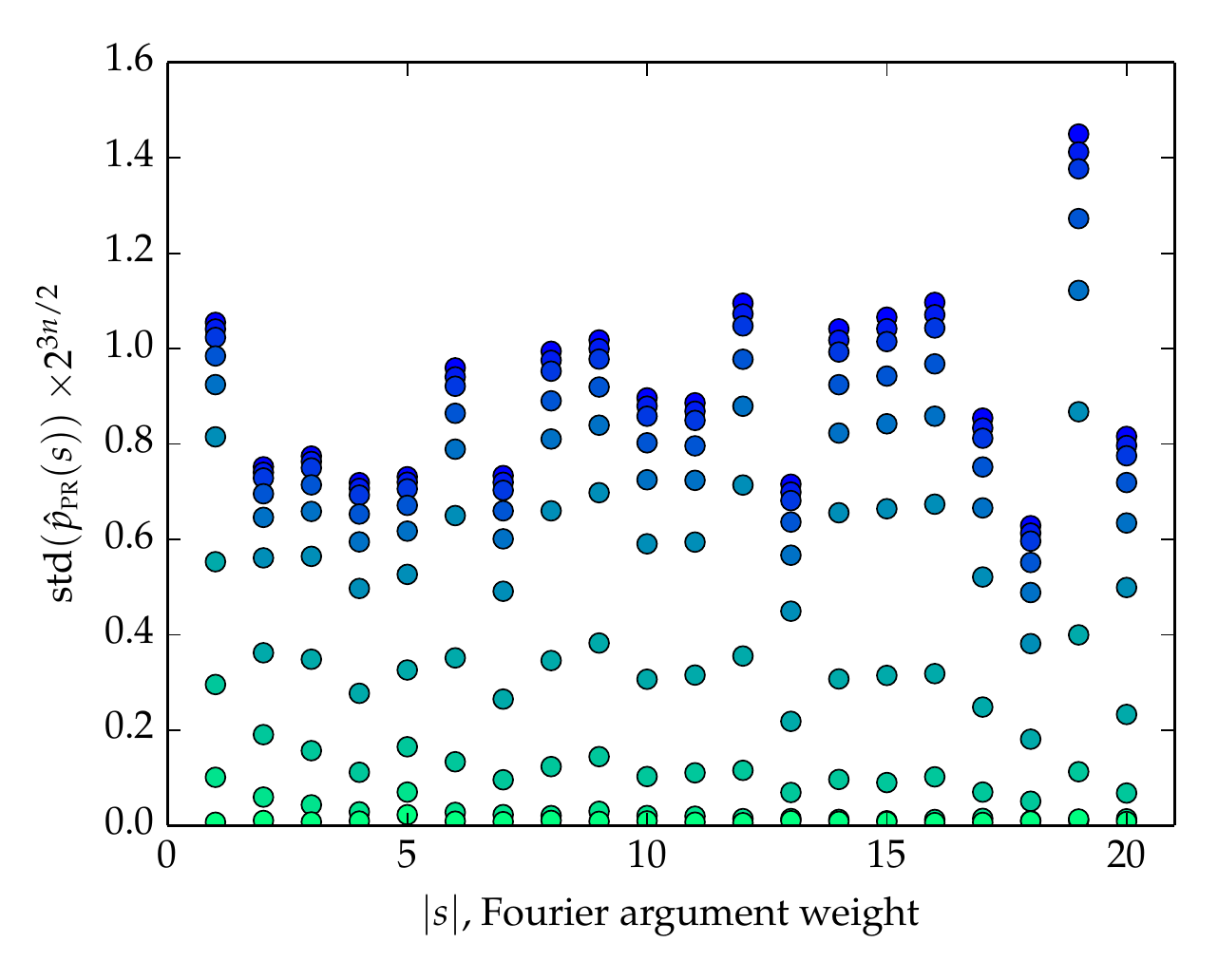}
  \caption{Numerical standard deviation of the Fourier components for the output distribution of pseudo-random universal quantum circuits~\cite{boixo2016characterizing}, as in Fig.~\ref{fig:fourier_component}, rescaled  by the standard deviation $2^{-3 n /2}$ from Porter-Thomas. Numerics confirm the Porter-Thomas estimates, and all standard deviations converge to 0 (uniform random distribution) with increased noise.  }
  \label{fig:fourier_std}
\end{figure}

Figure~\ref{fig:fourier_component} shows  Fourier components obtained numerically from the output distribution $\{p_\PR(x)\}$ of low-depth pseudo-random universal quantum circuits~\cite{boixo2016characterizing}, rescaled  by $2^{-3 n /2}$. Colors from blue to green correspond to increasing noise $\epsilon$. Figure~\ref{fig:fourier_std} shows the numerical standard deviation, also rescaled. The numerics confirm the Porter-Thomas estimates: all Fourier components are of order $\hat p_\PR(s) \in O(2^{-3n/2})$, and they converge to 0 (the random uniform distribution) with increasing $\epsilon$. Therefore, guessing the uniform random distribution is already a good approximation to all the Fourier components $2^n \hat p_\PR(s)$ with an additive error $O(2^{-n/2})$~\footnote{Nevertheless, note that there are an exponential number of Fourier components and that the statistical distance between the ideal and the uniform distribution is $\sum_x |p_\PT(x) - 1/N| = 2/e$.}. 

A critical subroutine of the polynomial approximation algorithm to sample from IQP circuits~\cite{bremner16} (and extended to pseudo-random circuits in Ref.~\cite{yung2017can}), is the approximation of a Fourier component $\hat p_\IQP(s)$ with additive error $\eta/2^{n}$ at cost $O(1/\eta^2)$. We just showed above, both numerically for pseudo-random circuits, and analytically for the Poter-Thomas distribution, that $2^n \hat p(s) \in O(2^{-n/2})$. Therefore, improving this trivial estimate for a single Fourier component requires $\eta \lesssim 2^{-n/2}$ and, according to the Chernoff bound, cost $O(2^n)$. Even for IQP circuits, using the Porter-Thomas approximation,  this is already of the same order as evaluating the probability exactly. When this technique is used for chaotic random circuits, the cost is exponential in $n' = m+n$. 

In conclusion, noisy pseudo-random universal quantum circuits, as well as noisy IQP circuits, converge to the uniform random distribution exponentially in the error rate $\epsilon$. Therefore, approximating an experimental noisy output distribution for large $n$ can be done in cost $O(n)$. The polynomial algorithm based on Fourier analysis~\cite{bremner16,yung2017can} does not output a distribution correlated with the ideal distribution, because it does not improve over guessing uniformly at random. Therefore, it does not represent an additional challenge over previous results suggesting the failure of classical devices in simulating noisy chaotic circuits with about 48 qubits and depth 40, as long as the error rate per gate is low ($\sim 0.3\%$). This extreme sensitivity of chaotic evolutions to noise allows experimentalists to approximate the global fidelity of complex evolutions using cross-entropy benchmarking, and important metric of experimental progress. 


\begin{acknowledgments}
  \label{sec: Acknowledgment}
 We would like to thank Michael J. Bremner, Edward Farhi and Ashley Montanaro for discussions. 
\end{acknowledgments}

\appendix

\section{Distribution of Fourier Components}\label{app:pt_fourier_distr}

We give here a direct derivation of the form of the distribution of the probability of Fourier components $\Pr_\F(\hat p_\PT)$ assuming that the probabilities $\{p_\PT(x)\}$ have a Porter-Thomas (exponential) distribution~\cite{porter1956fluctuations,brody_random-matrix_1981,mehta2004random}. 

For a given bit-string $s$, the value of the Fourier coefficient is
\begin{align}
  \hat p(s) = \frac 1 {N} \sum_x p(x) (-1)^{s \cdot x}= \frac 1 {N} \( 1 - 2 \sum_{x \in S} p(x)\)\;,
\end{align}
where $x \in S$ if and only if $s \cdot x$ is odd, and $N=2^n$. The size of $S$ is $|S| = 2^{n-1}$ (with $s \neq 0$). We first calculate the distribution
\begin{align}
  \Pr_\S(u) =  \left\langle \delta\bigg(\sum_{x \in S} p(x) - u\bigg) \right\rangle \;.
\end{align}

 Assume  a quantum state $\ket \psi$ sampled uniformly at random in Hilbert space. We write a generic state in the basis $\{\ket{x}\}$ as $\ket \psi = \sum_x (a_x + i b_x) \ket x$. Then 
 \begin{align}
   \Pr_\S(u) = {{\rm num}(u) \over {\rm dem}}
 \end{align}
where
\begin{multline}
  {\rm num (u)} = \int_{- \infty}^\infty \Pi_x d a_x d b_x \,\delta\bigg(\sum_x a_x^2 +b_x^2 -1\bigg) \\ \delta\bigg(\sum_{j=1}^{N/2} a_j^2 +b_j^2 -u\bigg) \;,
\end{multline}
and
\begin{align}
  {\rm dem} = \int_{- \infty}^\infty \Pi_x d a_x d b_x \, \delta\bigg(\sum_x a_x^2 +b_x^2 -1\bigg) \;.
\end{align}

We calculate the numerator as
\begin{align}
  &\int_{- \infty}^\infty \Pi_x d a_x d b_x \,\delta\bigg(\sum_x a_x^2 +b_x^2 -1\bigg) \delta\bigg(\sum_{j=1}^{N/2} a_j^2 +b_j^2 -u\bigg) \nonumber \\&=  \int_{- \infty}^\infty \Pi_x d a_x d b_x \,\frac 1 {2 \pi} \int_{- \infty}^\infty dt\,e^{i t \sum_x (a_x^2 + b_x^2) - i t}\nonumber \\&\quad\quad  \frac 1 {2 \pi} \int_{- \infty}^\infty dw\,e^{i w \sum_{j=1}^{N/2}\big(a_j ^2 + b_j^2\big) - i w u} \nonumber \\&=\frac 1 {2 \pi} \int_{- \infty}^\infty dt\, e^{-it}\( \int_{- \infty}^\infty  d a_x d b_x \,e^{i t (a_x^2 + b_x^2) } \)^{N/2} \nonumber \\&\quad\quad\ \frac 1 {2 \pi} \int_{- \infty}^\infty dw \,e^{-iwu}\( \int_{- \infty}^\infty  d a_x d b_x e^{i (t+w) (a_{x} ^2 + b_{x}^2) }\)^{N/2}\nonumber 
  \\ &=\frac 1 {2 \pi} \int_{- \infty+i\epsilon}^{\infty+i\epsilon} dt\,e^{-it} \(\frac \pi {-it}\)^{N/2}\nonumber \\&\quad\quad  \frac 1 {2 \pi} \int_{- \infty+i\varepsilon}^{\infty+i\varepsilon} dw\,e^{-iwu} \(\frac \pi {-i(t+w)}\)^{N/2} \nonumber 
  \\ &=- \frac {\pi^N}{(-i)^N} \frac 1 {(N/2-1)!} \lim_{t \to 0}  \frac {d^{N/2-1}}{dt^{N/2-1}} e^{-i t(1-u)} \nonumber \\&\quad\quad \frac 1 {(N/2-1)!} \lim_{w \to 0}  \frac {d^{N/2-1}}{dw^{N/2-1}} e^{-i w u} \nonumber \\ &= \pi^N \frac 1 {\((N/2-1)!\)^2}(1-u)^{N/2-1} u^{N/2-1}\;.
\end{align}

The denominator is calculated in a similar way
\begin{align}
  \int_{- \infty}^\infty &\Pi_x d a_x d b_x \,\delta(\sum_x a_x^2 +b_x^2 -1) \nonumber \\&=\frac 1 {2 \pi} \int_{- \infty}^\infty dt\, e^{-it}\( \int_{- \infty}^\infty  d a_x d b_x \,e^{i t (a_x^2 + b_x^2) } \)^{N} \nonumber \\&=\pi^N \frac 1 {(N-1)!}\;.
\end{align}

Therefore
\begin{align}
   \Pr_S(u) = \frac {(N-1)!} {\((N/2-1)!\)^2}(1-u)^{N/2-1} u^{N/2-1}\;.
\end{align}

For the distribution of a Fourier component
\begin{align}
   \hat p_\PT(s) = \frac 1 {N} \( 1 - 2 \sum_{x \in S} p(x)\) = \frac 1 {N} \( 1 - 2 u\)\;.
\end{align}
we have
\begin{align}
  \Pr_\F(\hat p_\PT) &= \frac N 2 \Pr_\S\((1-\hat p_\PT N)/2\) \\ &= \frac N 2 \frac {(N-1)!} {\((N/2-1)!\)^2}\(\frac 1 2 + \frac {\hat p_\PT N} 2\)^{N/2-1} \nonumber \\ &\quad\quad\quad\quad\quad \(\frac 1 2 - \frac {\hat p_\PT N} 2\)^{N/2-1}  \\ &= \frac {N^2} {2^{N+1}} { N \choose N/2} (1-\hat p_\PT^2 N^2) ^{N/2-1} \\ &\simeq \frac {N^{3/2}} {\sqrt {2 \pi}} \exp \( - \frac 1 2 \hat p_\PT^2 N^3\)\;.
\end{align}

In conclusion, the distribution of Fourier components is Gaussian with mean 0 and standard deviation $N^{-3/2}$.

\bibliographystyle{apsrev4-1} 
\bibliography{random_circuits}

\end{document}